\documentclass[twocolumn,showpacs,preprintnumbers,superscriptaddress,amsmath,amssymb]{revtex4-1}
\usepackage{graphicx,color,bm,natbib, placeins} 

\begin{document} 

\title{Doubly Transient Chaos: The 
Generic Form of Chaos in Autonomous Dissipative Systems}

\author{Adilson E. Motter} 
\affiliation{Department of Physics and Astronomy, Northwestern University, Evanston, IL 60208, USA}

\author{M\'arton Gruiz}
\affiliation{Institute for Theoretical Physics, E\"otv\"os University, P\'azm\'any P\'eter s\'et\'any 1/A, H-1117 Budapest, Hungary}

\author{Gy\"orgy K\'arolyi}
\affiliation{Institute of Nuclear Techniques, Budapest University of Technology and Economics, M\H{u}egyetem rkp. 9,  1111 Budapest, Hungary} 

\author{Tam\'as T\'el}
\affiliation{Institute for Theoretical Physics, E\"otv\"os University, P\'azm\'any P\'eter s\'et\'any 1/A, H-1117 Budapest, Hungary}
\affiliation{Theoretical Physics Research Group of the Hungarian Academy of Sciences at E\"otv\"os University, P\'azm\'any P\'eter s\'et\'any 1/A, H-1117 Budapest, Hungary}

\begin{abstract}
Chaos is an inherently dynamical phenomenon traditionally studied for trajectories that are either permanently erratic or transiently influenced by permanently erratic ones lying on a set of measure zero.  The latter gives rise to the final state sensitivity observed in connection with  fractal 
basin boundaries in {\it conservative} scattering systems and {\it driven} dissipative systems. 
 Here we focus on the most prevalent case of {\it undriven dissipative} systems, whose transient dynamics fall outside the scope of previous studies since no time-dependent solutions can exist for asymptotically long times. We show that  such systems can exhibit 
positive finite-time Lyapunov exponents and fractal-like basin boundaries which nevertheless have codimension one. In
sharp contrast with its driven and conservative counterparts, the settling  rate to the (fixed-point) attractors grows exponentially in time, meaning that the fraction of trajectories away from the attractors decays super-exponentially.  While no invariant chaotic sets exist in such cases, the irregular behavior is governed by {\it transient} interactions with  {\it transient} chaotic saddles,  which act as effective, time-varying 
chaotic sets.
\end{abstract}


\maketitle

As popularized by Gleick \cite{popularbook}, ``chaos is a science of process rather than state, of becoming rather than being.''   But the 
final state depends on the process and this has been widely explored in previous studies of transient chaos, where the object of analysis is not the (possibly simple) final behavior but instead the necessarily complicated transient dynamics leading to that outcome.  A canonical example is a periodically-forced damped pendulum with two periodic attractors and a fractal basin boundary separating them \cite{Grebogi}. 
A phenomenon of  continued research interest 
 \cite{Seoane2013,Pecora2012,Ravasz2011,Moura2011,WO,AE,HLKW},
transient chaos is determined by the presence of an invariant set that,  like in other manifestations of deterministic chaos, is formed by an uncountable number of aperiodic orbits that never settle down to periodic behavior and a dense set of unstable periodic ones \cite{Ott,chaosbook,LT}.  This invariant set is nonattracting and represents a zero-measure subset of the phase space whose stable manifolds form the fractal boundaries between regions  converging to different final states. It is thus the temporary approach to this chaotic saddle that gives typical orbits  transiently irregular dynamics, which in turn limits our ability to predict the final state. 

However important, these systems exclude a large and broadly significant  class of other systems that cannot have such an invariant set of time-dependent solutions. They are the dissipative but undriven 
(hence autonomous) systems that underlie numerous physical 
 processes  \cite{comment_aut}, including approach to thermodynamic equilibrium and various forms of self-organization and structure/pattern formation. Moreover, undriven dissipative systems exhibiting complex dynamics are common not only in physics, where a damped autonomous double pendulum is a prototypic example, but also in areas as diverse as chemistry, fluid dynamics, and astrophysics.

\begin{figure}[t!]
  \begin{center}
  \includegraphics[width=.48\textwidth ]{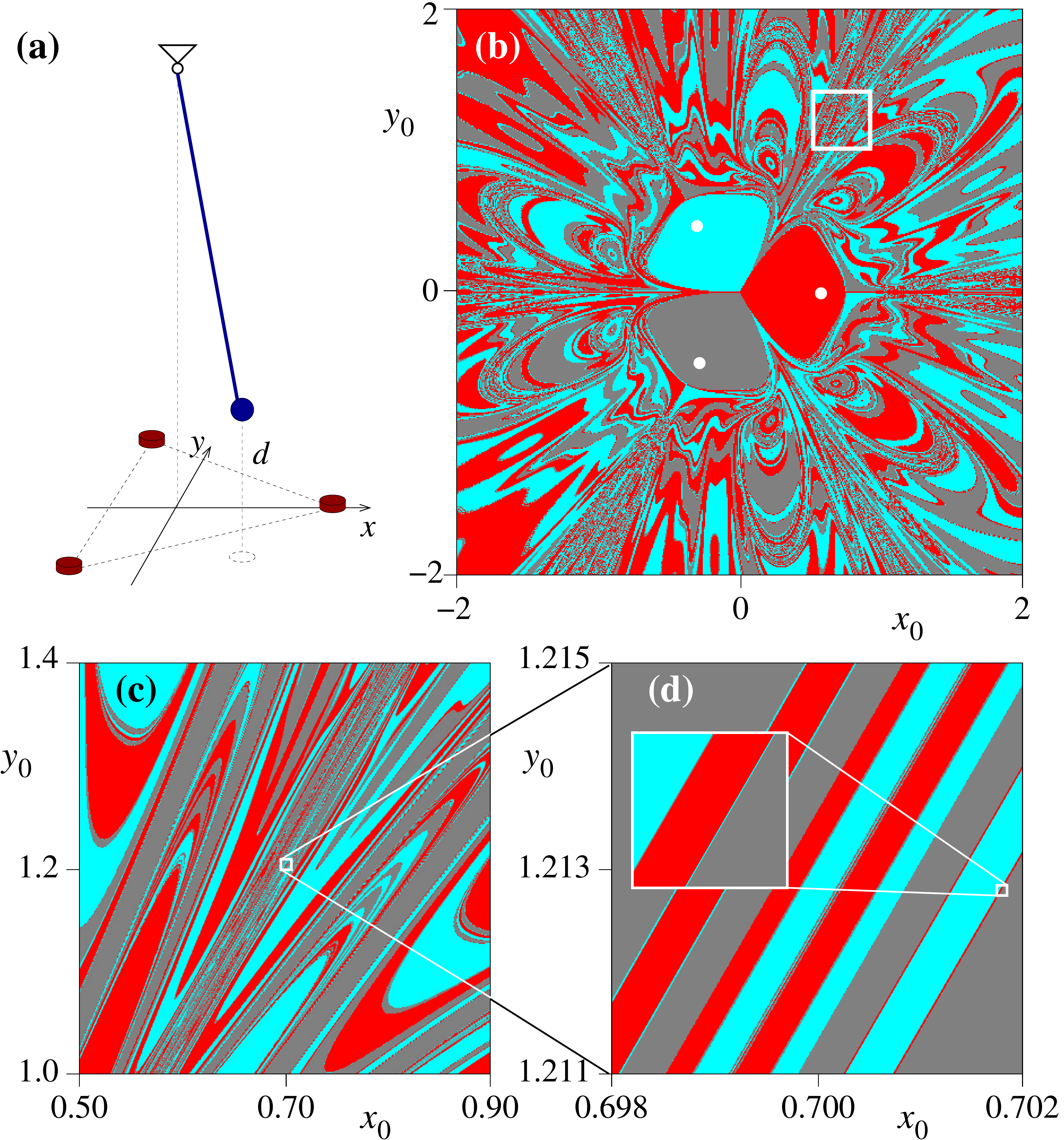}
    \vspace{-0.4cm}
  \end{center}
  \caption{(a) Autonomous magnetic pendulum as described in the text. 
                (b) Color-coded basins of attraction of the three fixed-point attractors of the system (white dots) for trajectories initiated with zero velocity.
                (c)-(d) Successive magnifications of the attraction basins shown in~(b).}
      \vspace{-0.2cm}
  \label{fig:magn}
\end{figure}

In  undriven physical systems  subject to
nonvanishing dissipation, the energy can only
decrease. As a result,  
the long-time behavior is necessarily very simple: each trajectory converges to one (out of possibly many) fixed point(s) 
in the case of the closed systems considered here.  
More important, this behavior is guaranteed for all orbits,  
not only for typical ones, indicating that typical orbits cannot experience the temporary influence of permanently chaotic ones. Yet, the dynamics can be  very complex for a transient period of time and the basin boundaries can be very intricate---properties that have often been associated with the concepts of transient chaos and fractals  \cite{chaosbook,peitgen}.
These are in fact  related to
the properties that give rise to the random-like behavior of coin tossing and die throwing   \cite{cointoss,gambling}.
Figure~\ref{fig:magn} shows the example of a magnetic pendulum with three fixed-point attractors, where the different colors mark the initial conditions associated with the different attractors.  Magnifications seem to reveal intermingled structures at smaller and smaller scales, which is suggestive of fractal basin boundaries and sensitive dependence on initial conditions. But can the
boundaries be fractal and the dynamics be transiently chaotic even though {\it all} motion eventually ceases?

In this Letter, we investigate the nature of the transient dynamics in undriven
dissipative dynamical
systems. We show that, due to the lack of long-time motion,  the behavior is of a completely different type compared to the one previously established for driven systems.  Our principal results are that in 
undriven systems: 
(i)  the measured dimension of the basin boundaries can be noninteger
and the finite-time Lyapunov exponents can be positive over all finite
scales but neither holds true asymptotically;  
(ii)  the basin boundaries have (asymptotic) fractal codimension one;
(iii) 
the survival probability of trajectories away from the attractors decays  super-exponentially, as
\vspace{-0.2cm}
\begin{equation}
P(t)\sim e^{-\frac{\kappa_0}{\gamma}e^{\gamma t}}, 
\label{eq1}
\end{equation}
leading to a settling 
 rate $\kappa(t) =\kappa_0e^{\gamma t}$, which grows exponentially in time;
(iv) while no invariant chaotic set exists on which long-time averages required for chaos characteristics can be defined, the transient behavior is governed by a transient chaotic saddle that is  prominent over a finite energy interval. We refer to this phenomenon as {\it doubly transient chaos}.

For concreteness, we focus on the magnetic pendulum as a model system, which, as we argue below, captures the generic properties of interest. The system consists of three identical magnets at the corners of a regular horizontal triangle of unit edge length and the pendulum itself,  formed by an iron bob suspended from above the center of the triangle through a massless rod  [Fig.~\ref{fig:magn}(a)]. The bob is subject to the influence of gravity, attractive magnetic forces, and drag due to air friction. For simplicity, we further assume that the length of the pendulum rod is long compared to the distance between the magnets, which 
allows us to describe the dynamics in terms of the  $(x,y)$~coordinates of the plane  using 
a small-angle approximation.
Following Refs.~\cite{peitgen,chaosbook}, 
we assume an inverse-square law interaction between the bob and the magnets as if they were point magnetic charges. The resulting dimensionless equations of motion are
\begin{eqnarray}   \label{eq:goveq1}
  \ddot{x}&=&-\omega_0^2 x-\alpha \dot{x}+
    \sum\limits_{i=1}^3\frac{\tilde{x}_i-x}{D_i(\tilde{x}_i,\tilde{y}_i)^3},\\
  \ddot{y}&=&-\omega_0^2 y-\alpha \dot{y}+
    \sum\limits_{i=1}^3\frac{\tilde{y}_i-y}{D_i(\tilde{x}_i,\tilde{y}_i)^3},
  \label{eq:goveq2}
\end{eqnarray}
where $(\tilde{x}_i,\tilde{y}_i)$ are the coordinates of the $i$th magnet, $\omega_0$ is the natural frequency, and $\alpha$ is the damping coefficient; 
here $D_i(\tilde{x}_i,\tilde{y}_i)=\sqrt{(\tilde{x}_i-x)^2+(\tilde{y}_i-y)^2+d^2}$ and $d$
are the distances from the pendulum bob to the  $i$th magnet and to the magnets' plane, respectively.
The coordinates of the magnets are $(\tilde{x}_1,\tilde{y}_1)=(\frac{1}{\sqrt{3}}, 0)$,
$(\tilde{x}_2,\tilde{y}_2)=(-\frac{1}{2\sqrt{3}}, -\frac{1}{2})$, and $(\tilde{x}_3,\tilde{y}_3)=(-\frac{1}{2\sqrt{3}}, \frac{1}{2})$.
In our simulations we set $\omega_0=0.5$, $\alpha=0.2$, and $d=0.3$ 
(except when
stated otherwise), which is representative of all cases for which the fixed point at the origin is unstable.
The magnetic pendulum then has three stable fixed points that serve as attractors for the long-time dynamics, as anticipated above and
shown in Fig.~\ref{fig:magn}(b)
for the bob released from positions $(x_0,y_0)$ with zero initial velocity.

\begin{figure}[t!]
  \begin{center}
  \includegraphics[width=.48\textwidth ]{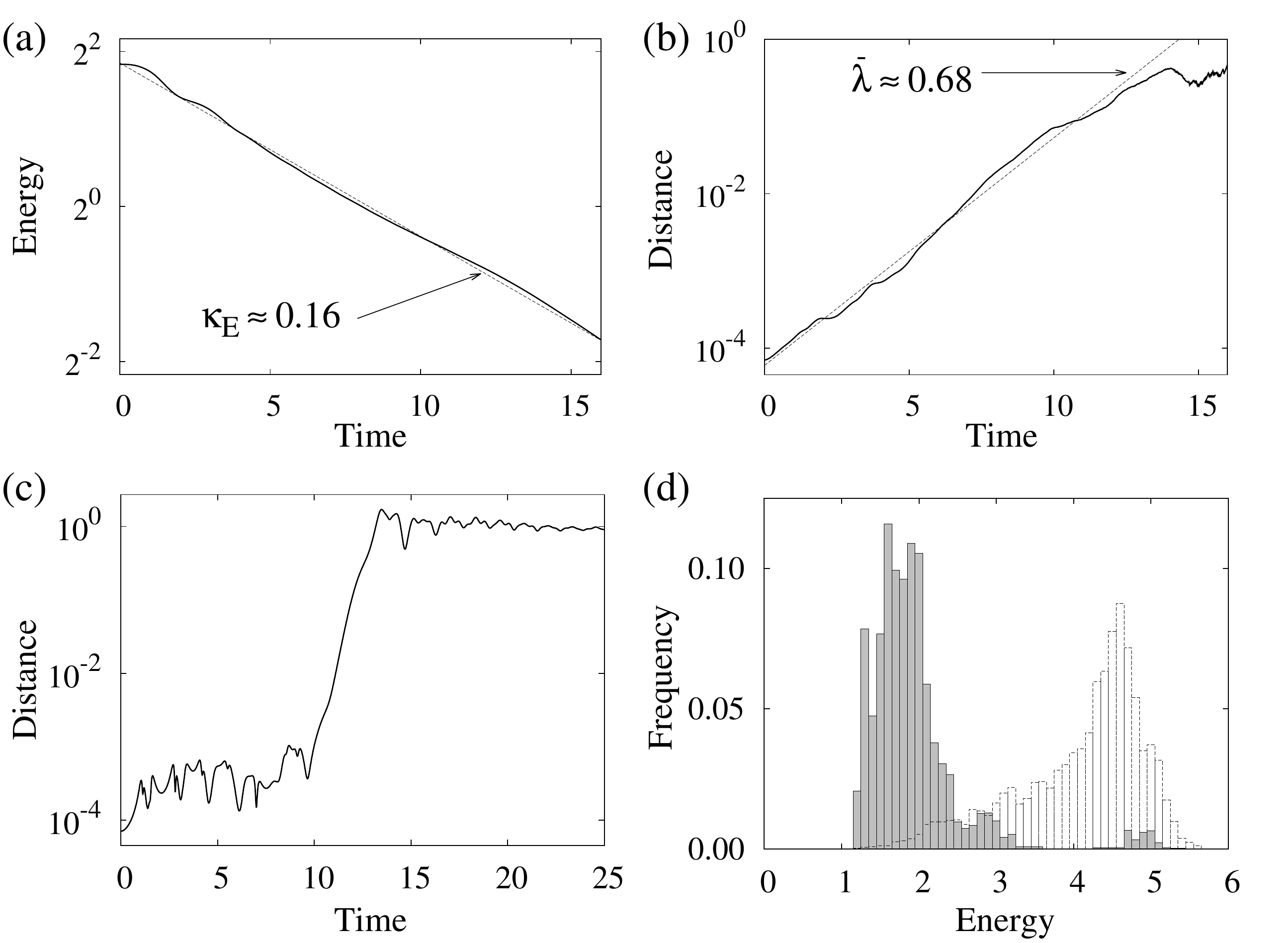}
    \vspace{-0.4cm}
  \end{center}
  \caption{
(a)  Average rate of energy dissipation 
and (b) average finite-time Lyapunov exponent, 
both estimated from 20,000 randomly selected initially 
close
pairs of trajectories belonging to different basins.
(c)  Time evolution of the distance between one such pair, showing the period of exponential separation.
(d) Distribution of the average energy of the trajectories during the exponential-separation phase. 
The background histogram shows the initial energy distribution. 
In (b), the time and Lyapunov exponent itself are measured over the period of exponential separation.
} 
     \vspace{-0.2cm}
  \label{fig:endec}
\end{figure}

First consider the average rate  $\kappa_E$ of energy dissipation due to damping. The
energy decays as $E(t)\sim \exp(-\kappa_E t)$ with $\kappa_E\approx 0.16$ [Fig.~\ref{fig:endec}(a)]. This 
average dissipation rate is close to the damping  coefficient $\alpha$  in Eqs.~(\ref{eq:goveq1})-(\ref{eq:goveq2}), and for
long enough times we indeed find $\kappa_E\to \alpha$, 
 as expected for 2D pendulum oscillations around a stable equilibrium.
Two nearby trajectories  in different basins tend to separate from each other over a relatively short period of time but they do so exponentially fast,  as 
illustrated in Fig.~\ref{fig:endec}(c). During the period of exponential separation, a small initial distance
 $\delta$ diverges as $\delta \exp(\overline{\lambda}t)$,
where $\overline{\lambda}\approx 0.68$ is the average finite-time (largest) Lyapunov exponent, 
which is approximately constant over a relatively long time for the aggregate of trajectories close to the basin boundaries [Fig.~\ref{fig:endec}(b)].
The average 
 energies of the trajectories during exponential separation fall within a 
narrower range 
than the initial energies
 [Fig.~\ref{fig:endec}(d)], indicating that neighboring high-energy trajectories tend to move together whereas low-energy ones have already settled into their attractors. 
 The deviation of the average dissipation rate $\kappa_E$ from $\alpha$ during the period of exponential separation indicates that fast separation takes place when the speed of the pendulum is low, as it would be expected when an orbit approaches an unstable fixed point or chaotic saddle. 
 
\begin{figure}[t!]
  \begin{center}
 \includegraphics[width=.45\textwidth ]{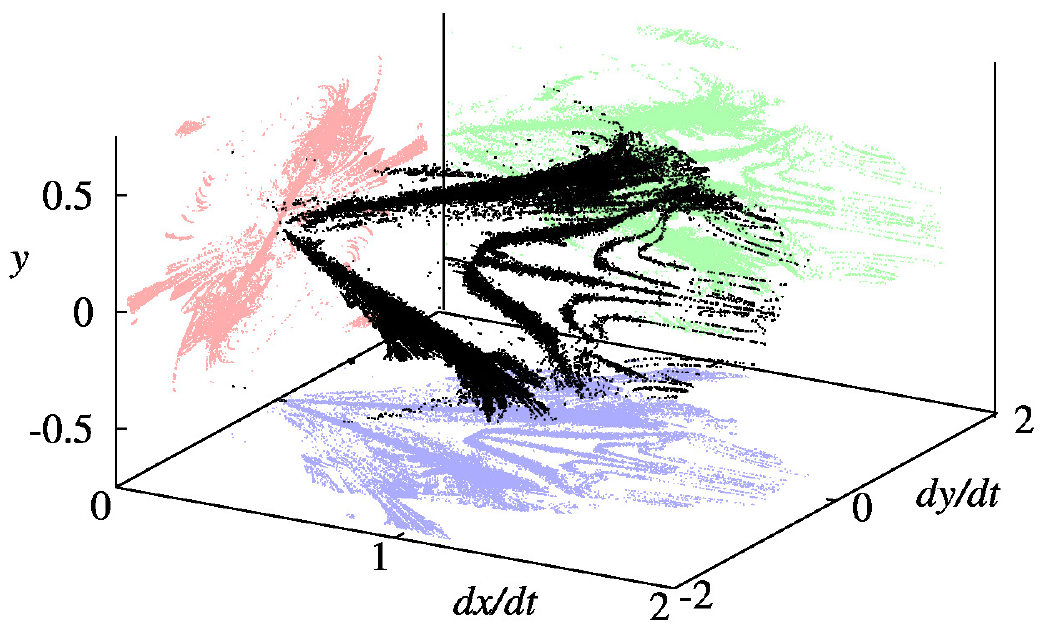}
   \vspace{-0.4cm}
  \end{center}
  \caption{Transient chaotic saddle of the magnetic pendulum~(black), represented through the 
  Poincar\'e section defined by $x=0$ and $\dot{x}>0$. The colored shades
  correspond to projections
  of the set on different coordinate planes. 
  }
     \vspace{-0.2cm}
  \label{fig:set}
\end{figure}

Our system does not have a chaotic saddle; it has in fact only a handful of unstable periodic orbits and all of them are fixed points. These include the fixed point at the origin, three along the symmetry axis connecting the attractors to the origin, and possibly a few others for specific parameter choices.
Boundary points between the different basins of attraction are expected to belong typically to the stable manifolds of fixed points 
that are locally stable along three directions and unstable along one direction in the 4D phase space. The unstable fixed point at the origin does not satisfy this condition since, for the parameters we consider, it has only two stable directions. The three unstable fixed points along the symmetry axes, however, have three eigenvalues with negative real parts.
We  argue, nevertheless, that this description
alone does not capture the complexity of the observed dynamical behavior.
Motivated by the apparent similarity  to  
transient chaotic dynamics, we propose that during the period of rapid separation the trajectories wander erratically in the vicinity of a set that plays the role of a chaotic saddle.  This set can be estimated from the positions where the  trajectories are while they separate exponentially from each other. The result is shown in Fig.~\ref{fig:set} and is strikingly similar to the usual chaotic saddles governing transient chaos. However, this set  consists of only pieces of trajectories in the phase space and as such is not an invariant set of orbits. Moreover, this set manifests itself only during the period of exponential separation,  which motivates us to refer to it as a {\em transient} chaotic saddle.

\begin{figure}[t!]
  \begin{center}
   \includegraphics[width=.48\textwidth ]{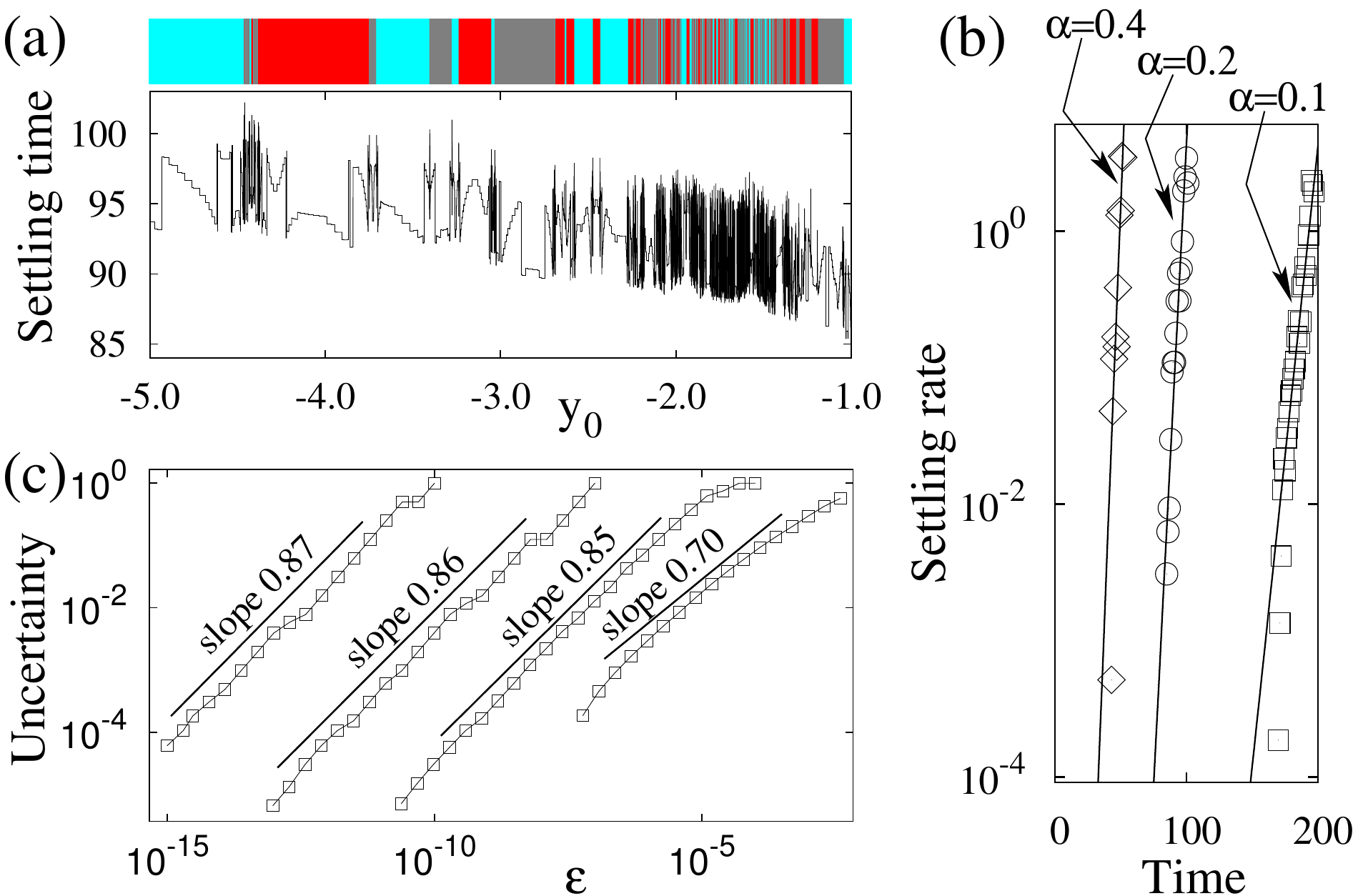}
    \vspace{-0.4cm}
  \end{center}
  \caption{(a) Settling time as a function of the initial $y$ coordinate for  trajectories initiated with zero velocity on the line  $x=-1$ to reach
a phase-space distance $10^{-4}$ from any of the attractors. The top bar indicates the corresponding basins of attraction, as color-coded in Fig.~\ref{fig:magn}.
 (b)~Settling rate $\kappa$ for different values of the damping coefficient $\alpha$, which increases exponentially as a function of time.
 (c) Estimation of the basin boundary (fractal) dimension using the uncertainty algorithm at successively smaller scale $\varepsilon$ along the line considered in panel (a).
  }
     \vspace{-0.2cm}
\label{fig:escaperate}
\end{figure}

A central aspect of dissipative systems concerns the time the trajectories take to reach (a predefined neighborhood of) any of the attractors,  which is referred to as the {\it settling time}  and is analogous to the escape time in open systems. Figure~\ref{fig:escaperate}(a) shows the settling time for trajectories of our system initiated on a straight line with zero initial velocity. 
This function exhibits a set of infinitely high peaks determined by the intersections of the initial line with the stable manifold of the nonattracting invariant set 
(which 
are typically basin boundary points). 
In a driven hyperbolic chaotic transient,
these singular points would form a Cantor set   
that is statistically self-similar.  
In our case the settling time is fundamentally different, exhibiting no self-similar structures. 
The singular points still form a 
set that resembles those of driven systems over several decades, but subsequent magnifications 
indicate
that this set (and hence the basin boundaries) become increasingly sparse at sufficiently small scales (see Supplement \cite{SI}, Fig.~S2).
Next, we quantify this systematic scale dependence.

Various dynamical quantities of a chaotic set can be determined from a single generating function---the  free energy function $F(\beta)$ \cite{beck,LT}. 
This function is defined as $\beta F(\beta)=-\lim_{t\to\infty} \frac{1}{t}\log I(\beta,t)$ for  $I(\beta,t)=\sum_{i=1}^{N(t)}\left(\ell_i(t)\right)^\beta$, where $N(t)$ is the number of intervals on a line of initial conditions (intersecting the stable manifold of the saddle) whose orbits have a settling time 
 larger than $t$, and  $\ell_i(t)$ are the lengths of these intervals. Quantities such as Lyapunov exponents, settling rates, dimensions, and entropies, which are by definition time independent and asymptotic,  can all be calculated directly from this function and its derivatives. In 
undriven dissipative systems, the $t\to\infty$ limit is
of little interest since all motion eventually ceases. But based on the settling time
 distribution of Fig.~\ref{fig:escaperate}(a), we can introduce a finite-scale free energy function as $\beta F(\beta,t)=-d\log I(\beta,t)/dt$. This function is now time dependent, which means that the  resulting dynamical quantities can be scale dependent. This dependence can be weak, as in the case of   the Lyapunov exponent that was found to be nearly constant over the range of scales, or very strong, as we demonstrate now for the settling rate.

We thus define the settling rate as the  instantaneous rate $\kappa(t)$ of decay of the fraction $P(t)$ of still unsettled trajectories at time $t$:  $dP(t)/dt=-\kappa(t)P(t)$. This corresponds to $\kappa(t)=\left.\beta F(\beta,t)\right|_{\beta=1}$ when expressed using the free energy function.  As shown in Fig.~\ref{fig:escaperate}(b), the instantaneous settling rate $\kappa(t)$ increases exponentially as a function of time, where the scaling exponent  is
$\gamma=0.21$, $0.43$, $0.56$ for $\alpha=0.1$, $0.2$, $0.4$, respectively. This represents a super-exponential decay  of $P(t)$, as summarized in Eq.~(\ref{eq1}),  which becomes increasingly more pronounced as the damping coefficient $\alpha$ is increased.  This is fundamentally different from the constant settling rate and power-law decay reported in the existing literature of  hyperbolic and nonhyperbolic transient chaos, respectively~\cite{Ott,chaosbook,LT}. 
An explanation for the super-exponential decay is  that (due to the exponential loss of energy) the difference between the settling times of two different trajectories scales with the difference of the logarithm of their initial energies, as $\Delta t\approx \frac{1}{\kappa_E}\Delta\ln E_0$, which causes them to reach the respective attractors after a comparable time.
While we used the average dissipation rate in this  
argument,
 note that the dissipation rate  of individual trajectories  is increasingly smaller for trajectories 
of same $E_0$
initiated increasingly closer to the basin boundaries.

The unbound, exponential increase of the settling rate has an important implication for the basin boundary:  its codimension is one. For an illustration, consider a single-scale Cantor set construction in which the proportion of the interval length removed at step $i$ is $\lambda_i$. At 
step $n$, there are $2^n$ intervals of length $\varepsilon_n=l_n/2^n$, where  $l_n=\Pi_{i=1}^{n}(1-\lambda_i)$. The box-counting dimension of the limit set then is 
$D_0=\lim_{n\rightarrow\infty} \frac{\ln 2}{\ln 2 -(\ln l_n)/n}$.  In a self-similar Cantor set, as often used to model hyperbolic chaotic systems, $\lambda_i=\lambda$  
(i.e., the fraction removed is the same for all $i$) and hence $(\ln l_n)/n=\ln(1-\lambda)$, leading to a dimension  $0<D_0<1$. 
An example of non-self-similar Cantor set, used to model nonhyperbolic chaotic systems \cite{lau91}, 
is the one for which $\lambda_i=1/(i+\lambda)$  (i.e., the fraction removed decreases with $i$) and hence $(\ln l_n)/n=\ln[\lambda/(n+\lambda)]/n$; this leads to $D_0=1$ even though the Lebesgue measure is zero.
The case of exponentially increasing settling rate corresponds to $l_n=e^{-\frac{\kappa_0}{\gamma} (e^{\gamma n} -1)}$ 
and hence $(\ln l_n)/n=-\kappa_0 (e^{\gamma n}-1)/\gamma n$. The dimension then is $D_0=0$ even though the set is uncountable.
But  since $\ln \varepsilon_n \sim -\frac{\kappa_0}{\gamma} e^{\gamma n}$ for large $n$, the convergence is in this case  logarithmically  slow with respect to the length scale, which requires going to very small scales for the accurate estimation of $D_0$ (as is also the case for other non-self-similar integer dimension Cantor sets); in this case, finite-scale calculations will always overestimate $D_0$. Similar result holds for any increasing settling rate such that $l_n= \lambda^{n^s}$ for $s>1$, which includes as particular case  $l_n=(\frac{2}{3})^{n^2}$, generated by taking $\lambda_i=1-(\frac{2}{3})^{2n-1}$.
Numerical calculation of the dimension of the basin boundaries in our system using the uncertainty algorithm 
\cite{greb1983}---which exploits  the scaling $\varepsilon^{\theta}$  of the fraction of points within a distance $\varepsilon$ of a basin boundary of codimension ${\theta}$---shows that the estimated ${\theta}$ becomes increasingly close to $1$ at smaller scales  (corresponding to basin boundaries of dimension $3$ in the full $4$-dimensional phase space) [Fig.~\ref{fig:escaperate}(c)].
This should {\it not} be taken as an indicator of minimal sensitive dependence on initial conditions, however, since sensitivity is minimal only when the asymptotic value of $\theta$ is approached, which, as suggested by our Cantor set construction and effectively demonstrated in Fig.~\ref{fig:escaperate}(c), requires extremely  small $\varepsilon$.

How general is the behavior described here?  
When driven by an external force, the magnetic pendulum exhibits the already known properties of driven dissipative systems (see Supplement  \cite{SI}). Thus the novel behavior identified here is indeed due to the autonomous nature of the dynamics as opposed to being  inherent to the magnetic pendulum itself.
As a rule of thumb, we suggest that systems that would be chaotic if the dissipation could be turned off are expected to exhibit doubly transient chaos for small but nonzero dissipation rates;
the dissipation rate sets the time scale
over which  trajectories will get intermingled by transient chaotic saddles.
In  particular, this is expected for Hamiltonian systems with mixed phase space, where the addition of dissipation generally converts the local minima of the energy at the center of Kolmogorov-Arnold-Moser islands into fixed-point attractors  \cite{contrast}.

For completeness, we contrast doubly transient chaos with other nonlinear phenomena in which signatures of chaos are observed in the absence of an invariant chaotic set. An important  case concerns strange nonchaotic  attractors  \cite{feudelbook} and repellors \cite{Moura2007},  which are dynamically-generated fractal invariant sets 
whose
largest Lyapunov exponents are nevertheless zero. Such behavior is usually induced by quasiperiodic driving, and hence refers to systems significantly different from those considered here. 
Another important case is stable chaos \cite{politi2010}, which is a spatio-temporal phenomenon in which the topological entropy can be positive even though the largest Lyapunov exponent is negative. Stable chaos  is usually studied in coupled map systems and  the phenomenon itself is rigorously observed in the thermodynamic limit.
Our characterization of undriven dissipative systems applies, however, to low-dimensional dynamics.

The doubly transient chaotic behavior analyzed here is both surprising and 
significant. 
Many authors  
have 
portrayed the dissipative magnetic pendulum and other such undriven systems in the same class as driven 
 dissipative systems, for the excellent reason that at first glance their basin boundaries and transient dynamics do seem similar. As shown here, however, they are fundamentally different and this is reflected both quantitatively and qualitatively. A remarkable distinction is that   
undriven dissipative
systems exhibit exponentially growing rather than constant settling rates
and, consequently,  fractal basin boundaries whose 
 complexity become increasingly diluted upon magnification.  
These 
properties are expected to be
common to many natural and man-made systems and, in particular, to those whose conservative counterpart is chaotic. 
The implications are thus rather general given the prevalence of chaos in 
conservative models 
and of
undriven
dissipative systems in the real world.
We suggest that our characterization of doubly transient chaos is relevant, for instance, for the study of
 ``transitional chaos"
in closed chemical reaction systems evolving toward thermodynamic equilibrium \cite{chem1,chem2}, 
in the
analysis of chaotic interacting vortices 
when dissipation due to viscosity is accounted for \cite{vortex1,vortex2}, 
and in the characterization of chaos in
spinning coalescing black hole-neutron star binaries and other binary  systems
as energy is lost due to gravitational waves \cite{gwave1,gwave2}.

The authors thank T. Nishikawa,
J.-R. Angilella, and V. Kalogera
 for illuminating discussions.
This research was supported by OTKA NK100296 and K100894, and NSF Grant PHY-1001198.

\newpage

\bigskip
\bigskip
\centerline{\Large Supplemental Material}
\vspace{0.7cm}

\renewcommand{\theequation}{S\arabic{equation}}

\setcounter{figure}{0}
\renewcommand{\thefigure}{{S\arabic{figure}}}

\noindent
\centerline{\bf Driven Dissipative Magnetic Pendulum}

\bigskip
It is instructive to compare the autonomous magnetic pendulum with a weakly driven one, which, as we show, does not exhibit super-exponential settling rates.  We  add time-dependence by moving the three magnets up and down sinusoidally, thus varying the vertical distance as $d=d_0+d_1\sin(\Omega t)$, which has the effect of not letting the trajectories rest in their final state.  We use $d_0=0.3$, $d_1=0.1$, and $\Omega=0.4$, which correspond to the same mean distance as before, a perturbation that leads to a small asymptotic kinetic energy,  and a driving frequency smaller 
than the natural frequency. Under these conditions, long-time motion converges to a persistent swinging of the pendulum in a vertical plane that, by symmetry, passes through the origin and 
one of the magnets. 

\begin{figure*}[h!]
  \begin{center}
 \hspace{-0.5cm} \includegraphics[width=0.70\textwidth ]{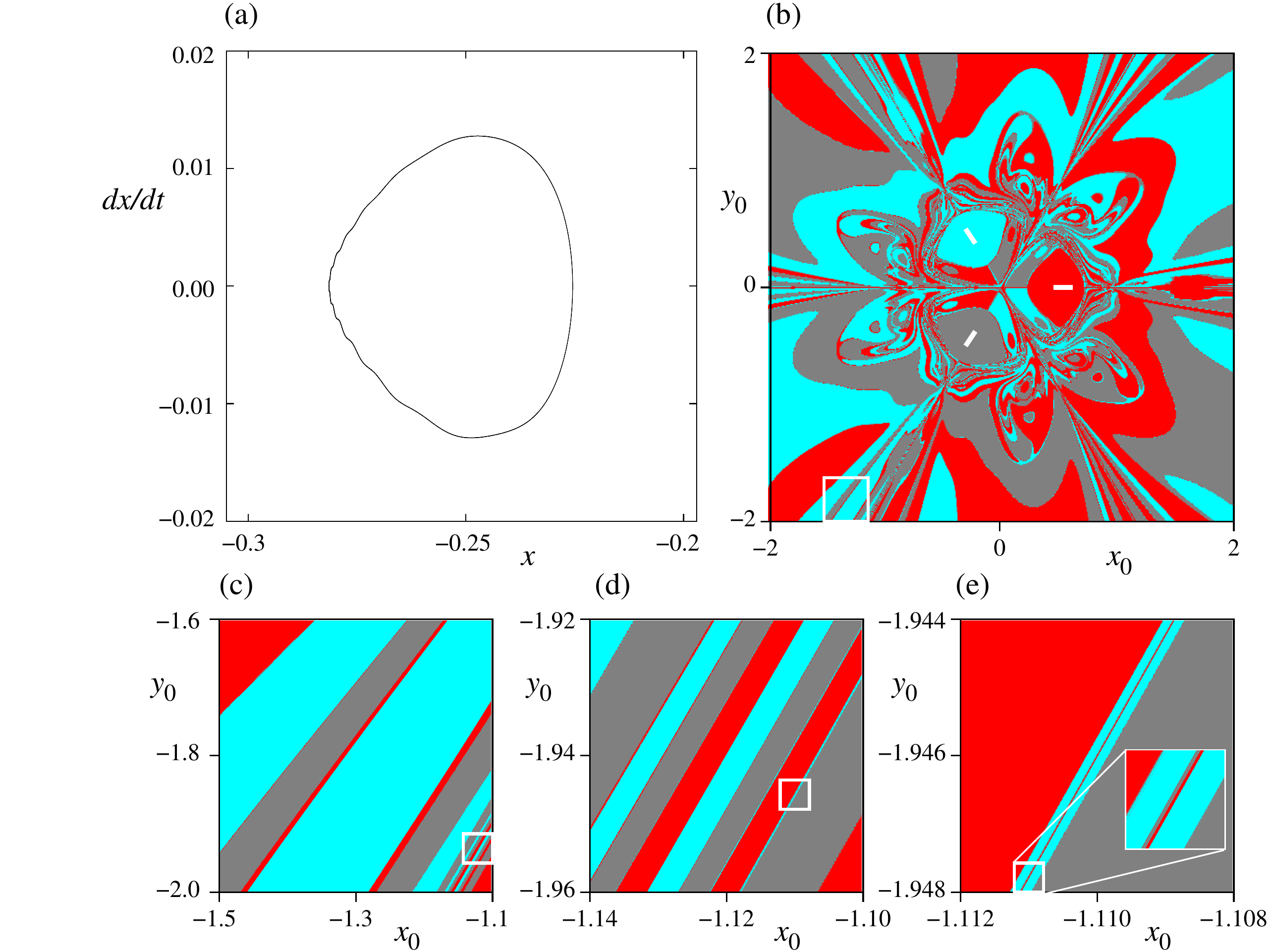}
               \vspace{-0.5cm}
  \end{center}
  \caption{
  Phase space of the driven magnetic pendulum.
  (a) Zoomed-in portrait of one of the three periodic attractors of the system in the $(x,\dot{x})$ projection of the space.
  (b) Color-coded attraction basins of the three periodic attractors  (line segments) for trajectories initiated with zero velocity, as in Fig.~1(b) (main text). 
  (c)-(e) Successive magnifications of the attraction basins. 
  }
  \label{figS1} 
\end{figure*}

\begin{figure*}[h!]
  \begin{center}
    \includegraphics[width=0.75\textwidth ]{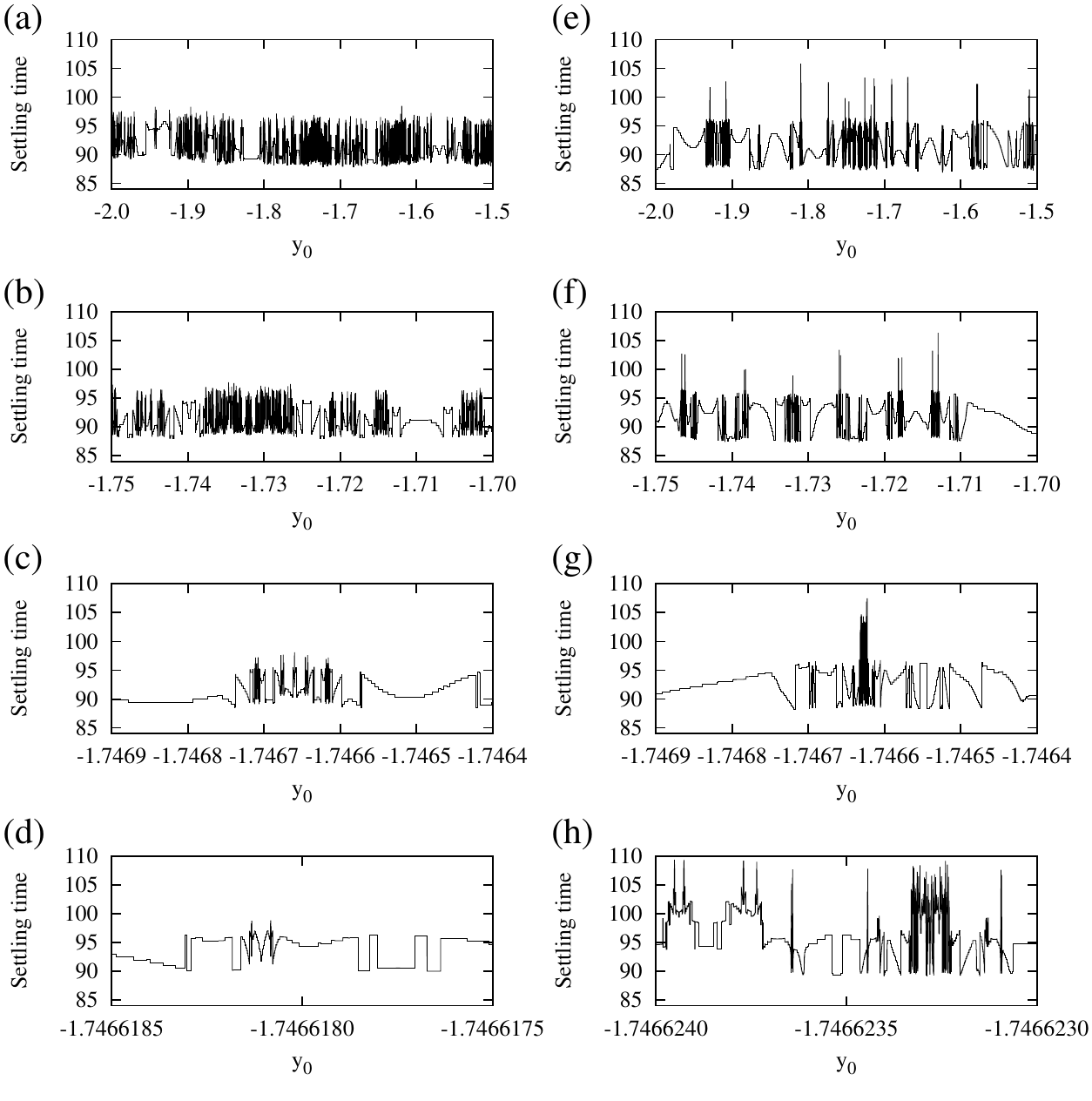} 
               \vspace{-0.5cm}
  \end{center}
  \caption{
  Settling time functions for  the (a)-(d) autonomous and (e)-(h) driven magnetic pendula.  
  The panels show successive magnifications 
  for trajectories initiated with zero velocity on the line $x=-1$  to
  reach a phase space distance $10^{-4}$ from an attractor. 
  Over orders of magnitude, the settling time of the autonomous system appears at least as complex as that of the driven system. Only at very small scales it becomes clear that the set of diverging settling times become increasingly sparse in the undriven case, as illustrated in panel (d),  while the driven system approaches an approximately self-similar structure at small scales. (The   nearly stepwise form of the settling time apparent in the bottom panels is an actual property of the systems rather than an artifact of resolution.)
  }
  \label{figS2} 
\end{figure*}

Therefore, the system now exhibits three periodic attractors  symmetrically disposed, each  close to a stable fixed point of the undriven pendulum---see Fig.~\ref{figS1}(a) for the $(x,\dot{x})$~projection of one of them. As shown in  Fig.~\ref{figS1}(b)-(e), the corresponding basins of attraction are at first glance very similar to those of the undriven case (cf.\ Fig.~1, main text)  in that they also consist of interwoven regular and complicated patterns.  At this level, the undriven and driven systems seem largely indistinguishable. Moreover, as shown in Fig.~\ref{figS2},  at large scales the settling time function of the undriven system seems as complex as that of the driven~one.

At very small scales, however, the distribution of diverging peaks of the settling time becomes increasingly sparse for the undriven system [Fig.~\ref{figS2}(d)] while systematic changes are far less pronounced for the driven case  [Fig.~\ref{figS2}(h)].  
The change over scales observed in the settling time of the undriven system is directly related to the slim fractal structure of the basin boundaries  
and is a reflex of the exponentially increasing settling rate
caused by energy dissipation.
Because we chose to drive the system only weakly to  facilitate comparison,
the driven system shows systematic changes 
at large  
scales, but only until the excess  energy is dissipated.

\begin{figure*}[h!]
  \begin{center}
    \includegraphics[width=.60\textwidth ]{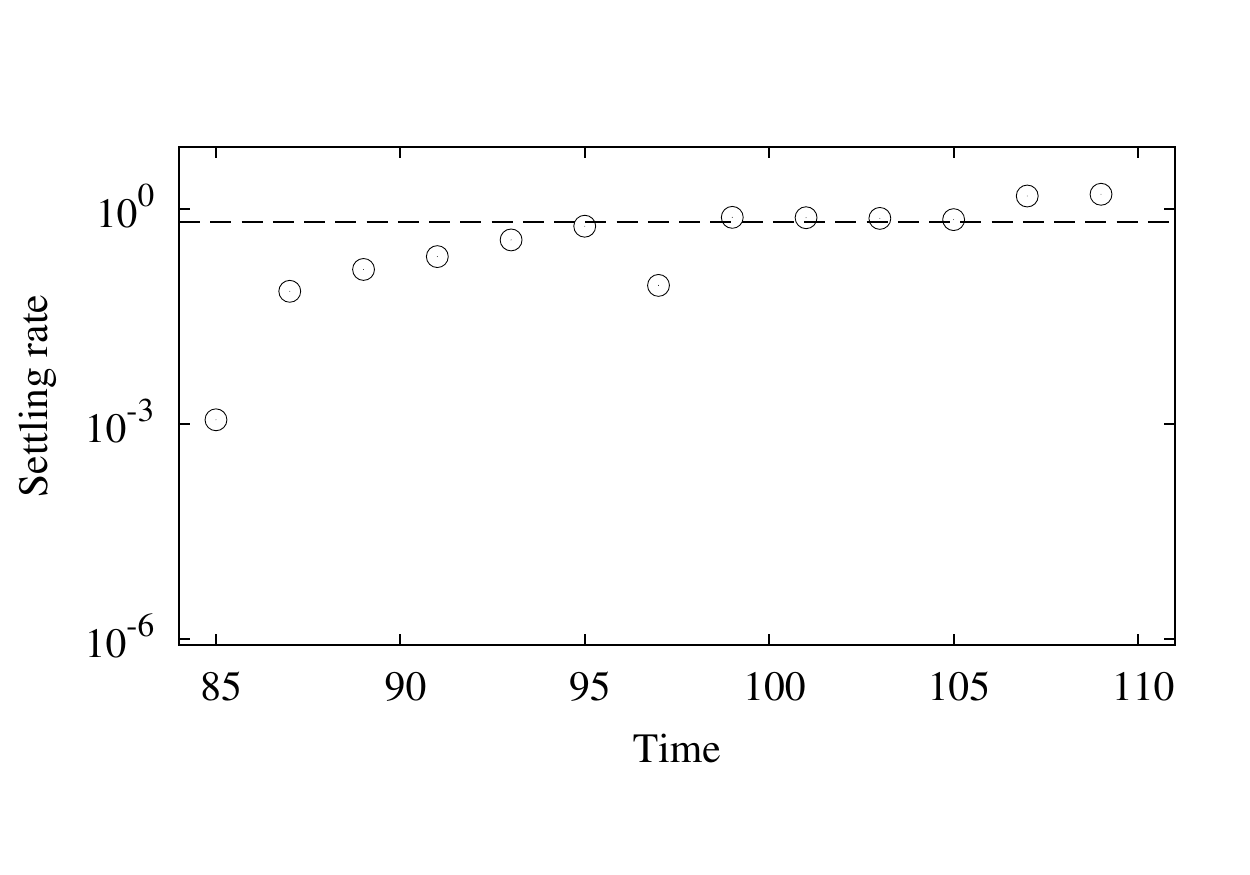}
               \vspace{-1.5cm}
  \end{center}
  \caption{
  Time dependence of the settling rate for the driven magnetic pendulum. The trajectories are initiated with zero velocity on the line  $x=-1$ 
  and are considered settled  once their phase-space distance to an attractor becomes smaller than $10^{-4}$.
  The circles represent the numerical results, while the dashed line is a guide to the eye.
  As for other driven dissipative systems, the settling rate in this case approaches a constant  rather than an exponentially growing function.
 }
  \label{figS3} 
\end{figure*}

It follows from the corresponding time-dependent free energy function  that the settling rate for the driven pendulum tends to a constant rather than an exponentially growing function (Fig.~\ref{figS3}). Again, the scale-dependence observed for  small settling times is due to the system being only weakly driven.
Thus, the properties of the driven pendulum are similar to  those of other such driven dissipative systems previously considered in the literature. On the other hand, the undriven pendulum considered in the main text has fundamentally different properties.


\begin{thebibliography}{99}
\bibitem{comment_aut}
We may use the term autonomous to refer to undriven when the context is clear,
but note that not all time-independent systems are undriven (e.g., the Lorenz system).
\end{thebibliography}
\end{document}